\documentclass[aps,prl,superscriptaddress,preprintnumbers,
showpacs,legalpaper,twoside,twocolumn]{revtex4}

\usepackage{bm}
\usepackage{graphicx}
\usepackage{amsfonts}
\usepackage{amsmath}
\usepackage{amssymb}

\begin{document}

\title{Quantum percolation in two-dimensional antiferromagnets}

\author{Rong Yu}
\affiliation{Department of Physics and Astronomy, University of Southern
California, Los Angeles, CA 90089-0484}
\affiliation{Kavli Institute for Theoretical Physics, University of California,
Santa Barbara, CA 93106-4030}
\author{Tommaso Roscilde}
\affiliation{Department of Physics and Astronomy, University of Southern
California, Los Angeles, CA 90089-0484}
\author{Stephan Haas}
\affiliation{Department of Physics and Astronomy, University of Southern
California, Los Angeles, CA 90089-0484}


\begin{abstract}
The interplay of geometric randomness and strong quantum
fluctuations is an exciting topic in 
quantum many-body physics, leading to the 
emergence of novel quantum phases in strongly
correlated electron systems. Recent investigations
have focused on the case of homogeneous site and 
bond dilution in the quantum
antiferromagnet on the square lattice, reporting
a classical geometric percolation
transition between magnetic order and disorder.  
In this study we show how  
{\it inhomogeneous} bond dilution leads  
to percolative quantum phase transitions, which we have
studied extensively by quantum Monte Carlo simulations.
Quantum percolation introduces a new class
of two-dimensional spin liquids, characterized 
by an infinite percolating network 
with vanishing antiferromagnetic order parameter.
\end{abstract}

\maketitle
Percolation occurs in a variety of
contexts, \cite{deGennes76} ranging from blood vessel
formation to clusters of atoms
deposited on substrate surfaces. 
A fundamental question is whether
the classical picture of permeating networks \cite{StaufferA94} 
applies as well to strongly fluctuating quantum systems. 
In this respect, low-dimensional quantum magnets offer
an ideal playground to probe the interplay between  
quantum fluctuations and geometric randomness both
theoretically and experimentally \cite{Vajketal02,Sandvik02}.

\begin{figure}[h]
\begin{center}
\null \hspace{-.9cm}
\includegraphics[bbllx=0pt,bblly=340pt,bburx=1100pt,bbury=940pt,%
     width=92mm,angle=0]{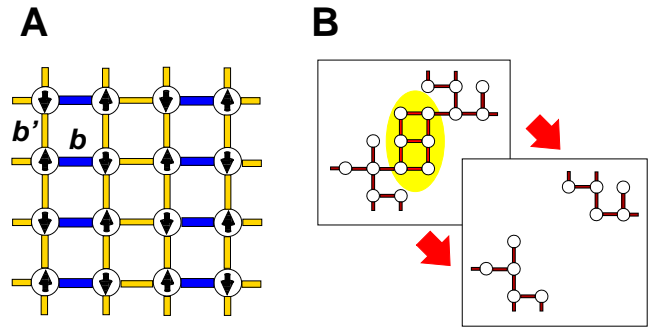} 
\caption{\label{structures} ({\it color online}) 
(A) A square lattice
quantum antiferromagnet composed of dimers (blue bonds, b)
and ladders (yellow bonds, b'). (B) A segment of ladder
in the percolating cluster
leads to local RVB states (yellow ellipsis). 
The spins involved in the RVB state
are very weakly correlated to the rest of the cluster,
causing an effective fragmentation of 
the percolating cluster into non-percolating subclusters.}
\end{center}
\end{figure}

Two electronic spins coupled by 
an antiferromagnetic Heisenberg interaction,
${\cal H} =  J {\bm S}_1\cdot {\bm S}_2$  
($J>0$), combine into a singlet state with 
purely quantum character.
Such singlets are the fundamental building blocks of many 
novel phases in quantum antiferromagnets. 
For instance, the ground state of a Heisenberg ladder
is a coherent superposition of singlet coverings
involving mostly nearest-neighbor spins. 
This {\it resonating valence bond} (RVB) state
\cite{Whiteetal94} is 
realized in a number of 
antiferromagnetically correlated materials for which magnetometry
and neutron scattering measurements indicate the absence
of magnetic order down to the lowest accessible temperatures
\cite{DagottoR96}. In contrast,  
the $S=1/2$ Heisenberg antiferromagnet on the square lattice
has an antiferromagnetically long-range-ordered ground state 
\cite{Manousakis91}. However, as shown in 
Fig. \ref{structures}(A), a square lattice can be decomposed 
into dimers and ladders in such a way that there are no two 
adjacent dimers/ladders. This suggests that, if lattice 
disorder can isolate some of these 
substructures, it may lead to quantum disordered ground states.

Recently, the effect of site dilution on
the square-lattice $S=1/2$ quantum Heisenberg 
antiferromagnet has been probed experimentally
in the compound La$_2$Cu$_{1-p}$(Zn,Mg)$_p$0$_4$,
in which magnetic Cu$^{2+}$ ions
are replaced randomly by non-magnetic
Zn$^{2+}$ or Mg$^{2+}$ ions \cite{Vajketal02}. 
The pure system ($p=1$) displays
long-range antiferromagnetic order 
\cite{Manousakis91}. 
Upon doping with static non-magnetic impurities, the order
parameter is reduced until it vanishes at 
a critical doping concentration $p_c$. Both experimental 
data \cite{Vajketal02} and numerical simulations  
\cite{Vajketal02,Sandvik02} indicate a magnetic
transition at the geometric percolation
threshold $p^{*}=0.40725$ \cite{StaufferA94} beyond which
the system breaks up into finite clusters. An analogous
result holds for the case of bond dilution, where a fraction
$p$ of bonds is cut at random in the system: again 
the onset of disorder is observed at the percolation
threshold $p_c=p^{*}=0.5$ \cite{Sandvik02}. In both cases this 
implies that the magnetic transition is controlled by the
underlying geometric transition. 
  A completely different picture of the effect of
disorder might instead emerge
when resonating valence bonds dominate the ground state 
of the system. In fact, the 
spins involved in a local RVB/singlet state are weakly correlated
to the remainder of the system, so that, from the point of view
of magnetic correlations, we can roughly consider these
spins missing from the geometric cluster they belong to, 
as sketched in Fig.~\ref{structures}(B). 
This implies that the percolating cluster is effectively fragmented
by the formation of RVB/singlet states, and therefore 
the magnetic transition is hindered with respect
to the geometric transition, acquiring in turn the nature
of a quantum phase transition. Nonetheless, according
to this argument the magnetic quantum transition 
appears as a renormalized percolative transition, with 
an effective geometry of the clusters dictated by 
quantum effects. This scenario is referred to as 
{\it quantum percolation}.

In this work we investigate the occurrence of quantum
percolation in the $S=1/2$ 
quantum antiferromagnet on the square lattice
with inhomogeneous bond dilution, modeled
by the Hamiltonian
\begin{equation}
 {\cal H} = \sum_{b} J_b
 {\bm S}_{1,b}\cdot{\bm S}_{2,b}+ 
 \sum_{b'} J_{b'}
 {\bm S}_{1,b'}\cdot{\bm S}_{2,b'}.
 \label{e.hamilton}
 \end{equation} 
\noindent
Here $\bm S$ are $S=1/2$ spin operators, 
the bonds $b$ are dimer bonds in Fig. \ref{structures}(B), 
and the bonds $b'$ are the inter-dimer bonds.
$J_b$ , $J_{b'} > 0$ are random antiferromagnetic couplings
drawn from a bimodal distribution taking the two values $J$ (`on')
and $0$ (`off'), with different probabilities of 
the `on'-state, 
$p(J_b=J) = P$ and $p(J_{b'}=J) = P'$. 
In the special 
case $P=P'$ the homogeneous bond-dilution problem is recovered: 
the percolation threshold coincides with a magnetic order-disorder
transition
at the classical critical value $P_c=P'_c=0.5$ 
\cite{StaufferA94}. In contrast, for 
$P\neq P'$ one encounters {\it inhomogeneous} 
bond-dilution.
In particular, at $P=1$ the system is an array 
of randomly coupled dimers, and in the opposite limit
$P'=1$ the system is a set of randomly coupled ladders. 
\begin{figure}[h]
\includegraphics[bbllx=50pt,bblly=40pt,bburx=560pt,bbury=490pt,%
     width=70mm,angle=0]{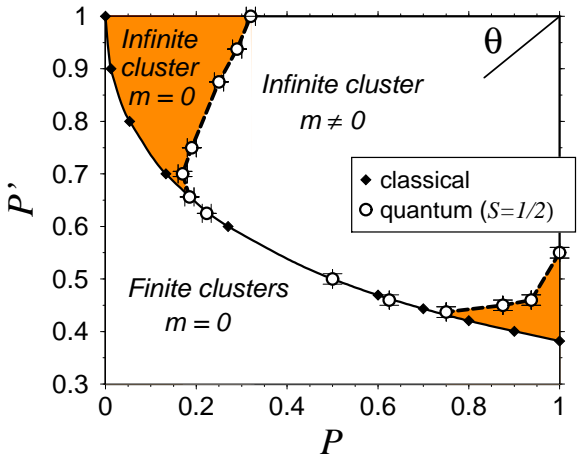}
\caption{\label{f.phdiagr} Phase diagram of the inhomogeneously
bond-diluted $S=1/2$ antiferromagnet on the square lattice.
The colored area indicates the quantum-disordered region
in which the system has developed an infinite percolating
cluster, but the magnetization $m$ of the system vanishes 
because of quantum fluctuations. The angle $\theta$ parametrizes 
the critical curve.} 
\vskip -.2cm
\end{figure} 

\begin{figure}[h]
\includegraphics[bbllx=14pt,bblly=80pt,bburx=590pt,bbury=500pt,%
     width=80mm,angle=0]{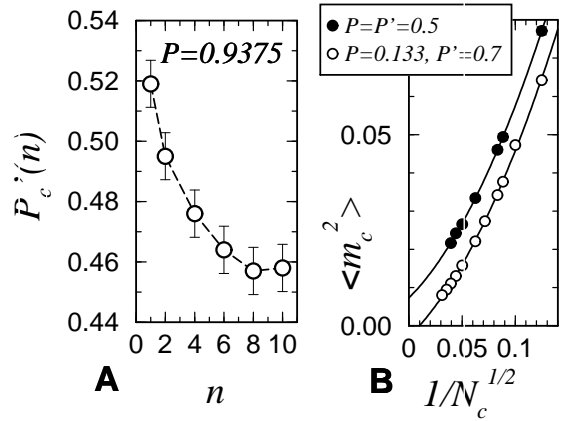}
\caption{\label{f.validation} (A) Estimates of the quantum critical point
as the value of the disorder showing the critical scaling $\xi_x,\xi_y \sim L$ 
at a temperature $T=J/nL$, plotted as a function of $n$.
For $n=8$ the estimate clearly becomes temperature-independent. 
(B) Disorder-averaged square order parameter on the percolating cluster only
as a function of the size $N_c$ of the percolating cluster 
for the homogeneous percolation point ($P=P'=0.5$) and for a classical 
inhomogeneous percolation point  ($P=0.133$, $P'=0.7$). The temperature
is $T=1/(8N_c)$. The solid lines are cubic polynomial fits.}
\vskip -.2cm
\end{figure}     
We determine the  geometric percolation
threshold for the inhomogeneous bond-dilution problem
by generalizing a very efficient algorithm recently developed 
for homogeneous percolation 
\cite{NewmanZ02}. The resulting phase boundary 
is shown with filled diamond symbols in Fig. \ref{f.phdiagr}. 
A continuous transition line, with constant critical 
exponents of 2D percolation \cite{StaufferA94}, is found, connecting
the percolation threshold for randomly coupled dimers,
${\bm P} = (P,P') = (1,0.382)$ with the 
percolation point ${\bm P}=(1/2,1/2)$ of the homogenously diluted
systems, and terminating at
the critical point for randomly coupled ladders,
${\bm P} = (0,1)$. In this limiting case an infinitesimal 
concentration of dimer bonds is sufficient to 
create a percolating cluster in the system.  

 When classical spins ($S=\infty$) are placed on each lattice site,
long-range antiferromagnetic order follows trivially 
the onset of geometric percolation. In contrast, 
this picture
changes drastically
for $S=1/2$ quantum spins. Approaching the 
two limits of randomly coupled dimers ($P \rightarrow 1$) and ladders
($P' \rightarrow 1$), the formation of singlet states for the 
spins on the dimers ($b$-bonds) or on the ladders 
($b'$-bonds) are 
favored statistically by geometry, such that quantum 
fluctuations are strongly enhanced. This contrasts
with the case of homogeneous bond percolation, $P=P'$,
in which quantum effects do not change the classical
picture \cite{Sandvik02}. Therefore the crossover from
homogeneous to inhomogeneous percolation in the $S=1/2$
antiferromagnet is accompanied by a continuous enhancement
of quantum fluctuations.

To determine the magnetic phase diagram
in the quantum limit we use
the Stochastic Series Expansion Quantum Monte Carlo (QMC) method
based on the directed-loop algorithm \cite{SyljuasenS02}.
This technique enables us to access ultra-low temperatures
on large $L\times L$ lattices whose size is
well inside the scaling range of the critical regime.
To obtain finite-size scaling at $T \rightarrow 0$ we have
carried out simulations on lattices as large as
$L = 64$. For strongly inhomogeneous systems, we have studied the
$T \rightarrow 0$ behavior by sitting at an inverse temperature 
$\beta J =8L$, at which the analysis of critical scaling leads
to a temperature-independent estimate of the 
transition point, as shown in Fig. \ref{f.validation}(A).
For systems closer to the homogeneous limit
we found necessary to cool the system to even lower 
temperatures, which we can efficiently reach by a 
$\beta$-doubling approach \cite{Sandvik02}.
Finite temperature properties are studied 
on larger clusters with $L = 200$.
Furthermore, simulations are also carried
out at the percolation threshold on percolating clusters of 
exact size $N_c$ grown freely from an initial seed 
(see Ref.\cite{Sandvik02} for a similar analysis).
For each value of $P$ and $P'$ we average
over 100$\div$300 different realizations
of the  bond disorder (disorder averages are indicated 
as $\langle...\rangle$).
The finite staggered magnetization $\sqrt{\langle m^2 \rangle} = 
\sqrt{3\langle S(\pi,\pi) \rangle/L^2}$
of an antiferromagnetically ordered ground state is accompanied
by a divergent spin-spin correlation length
along  the $x$ and $y$ direction, $\xi_x$
and $\xi_y$. The location of the magnetic transition is obtained from the
bond concentration at which the $T=0$ correlation lengths
diverge linearly with lattice size, $\xi_x,\xi_y \sim L$.

 The transition line for the quantum
antiferromagnet and for the classical
percolation problem are clearly different. This shows
that the magnetic phase transition in the case of inhomogeneous
bond dilution turns into a {\it quantum} phase transition
for sufficiently strong inhomogeneities, leaving
the percolation transition line at two \emph{multicritical} 
points. In particular,
between the two transition lines there is an intermediate 
regime (colored area) in which geometrical
percolation has been reached, such that there is
an infinite percolating cluster 
\cite{StaufferA94}. Nonetheless, magnetic
order is absent, as shown by the scaling of the
order parameter in Fig. \ref{f.validation}(B).
Here the disorder-averaged squared magnetization $m_c$ 
of the percolating cluster \emph{only} is shown as a 
function of the cluster size. 
Unlike the case of the homogeneous percolation point
\cite{Sandvik02} $P=P'=0.5$,  
the percolating cluster in the strongly inhomogeneous 
case shows a vanishing magnetization at the onset of 
geometric percolation, so that the addition of extra
bonds is required for the magnetic order to set in.
The {\it quantum disordered} regime 
of the phase diagram corresponds to an {\it infinite
family of two-dimensional quantum spin liquids}
for which the antiferromagnetic Heisenberg
interaction does not lead to spontaneous symmetry
breaking in the thermodynamic limit.

Moreover, the phase
diagram in Fig. \ref{f.phdiagr} shows remarkable reentrant
behavior. Decreasing $P$ ($P'$) from 1 to 0 with fixed $P'$ ($P$)
in the range $0.45 \lesssim P' \lesssim 0.55$
($0.2 \lesssim P \lesssim 0.32$), the system 
experiences first a magnetic transition from the spin
liquid to the ordered antiferromagnetic state. This 
is a clean example of the {\it order-by-disorder}
mechanism, in which magnetic order is induced
by randomness due to the geometric destruction of local 
RVB/singlet states.

 Unlike in other doped lattices \cite{orderbydisorder}, 
 we do not observe the occurrence 
of residual magnetic order due to the long-range effective 
interactions \cite{SigristF96}
between spins not involved in a local RVB/singlet state.
We ascribe this finding to a distinctive feature of bond
dilution as opposed to other forms of disorder: in fact,
when the cut of a bond does not lead to the local enhancement
of quantum fluctuations, it leaves always {\it two}
nearby spins not forming a singlet with their neighbors,
so that each unpaired spin finds immediately its partner
to build a longer-range singlet. 
\begin{figure}[h]
\includegraphics[bbllx=20pt,bblly=82pt,bburx=550pt,bbury=465pt,%
     width=65mm,angle=0]{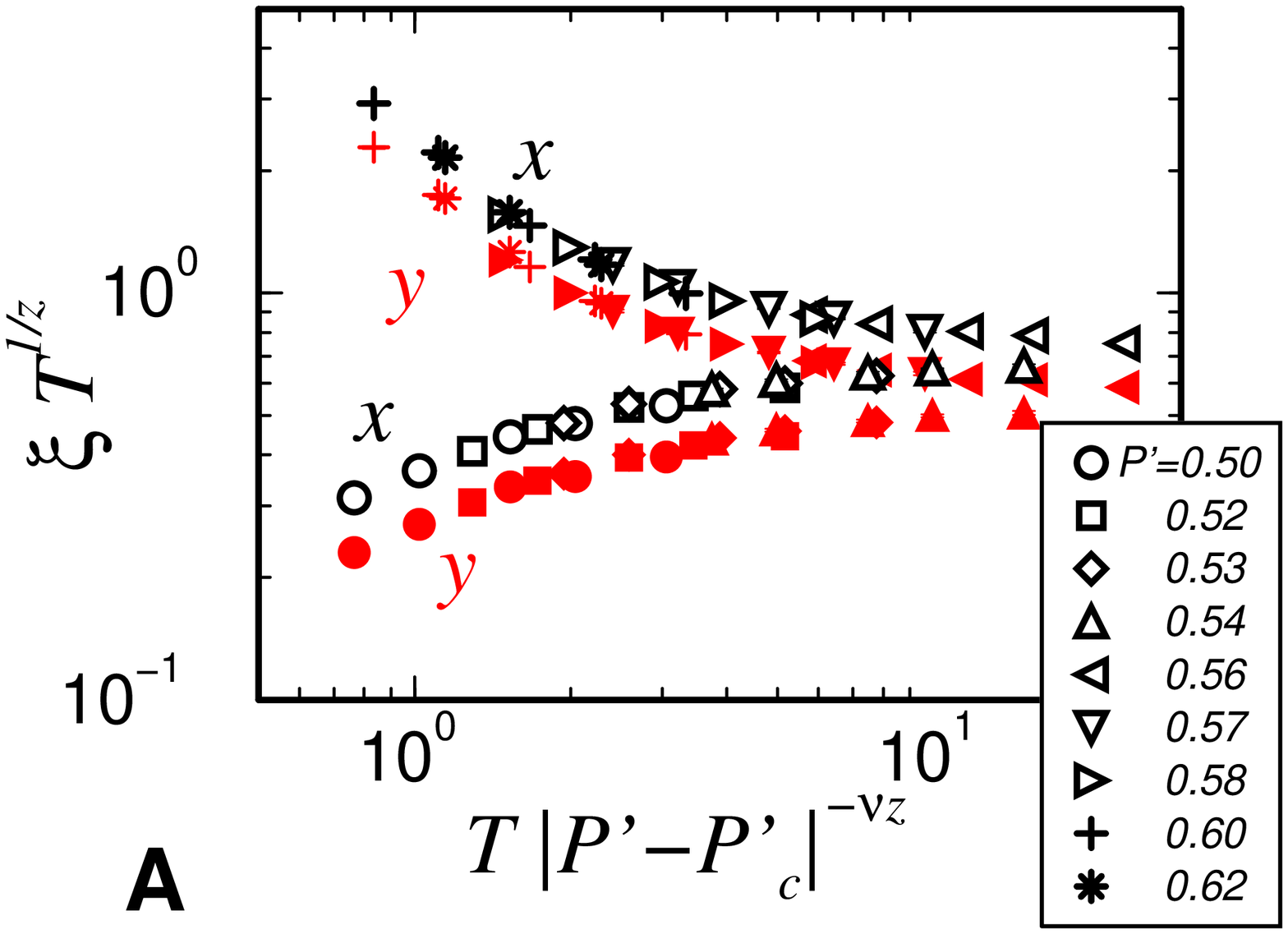}     
\null~~~~~~~\includegraphics[bbllx=10pt,bblly=82pt,bburx=550pt,bbury=490pt,%
     width=65mm,angle=0]{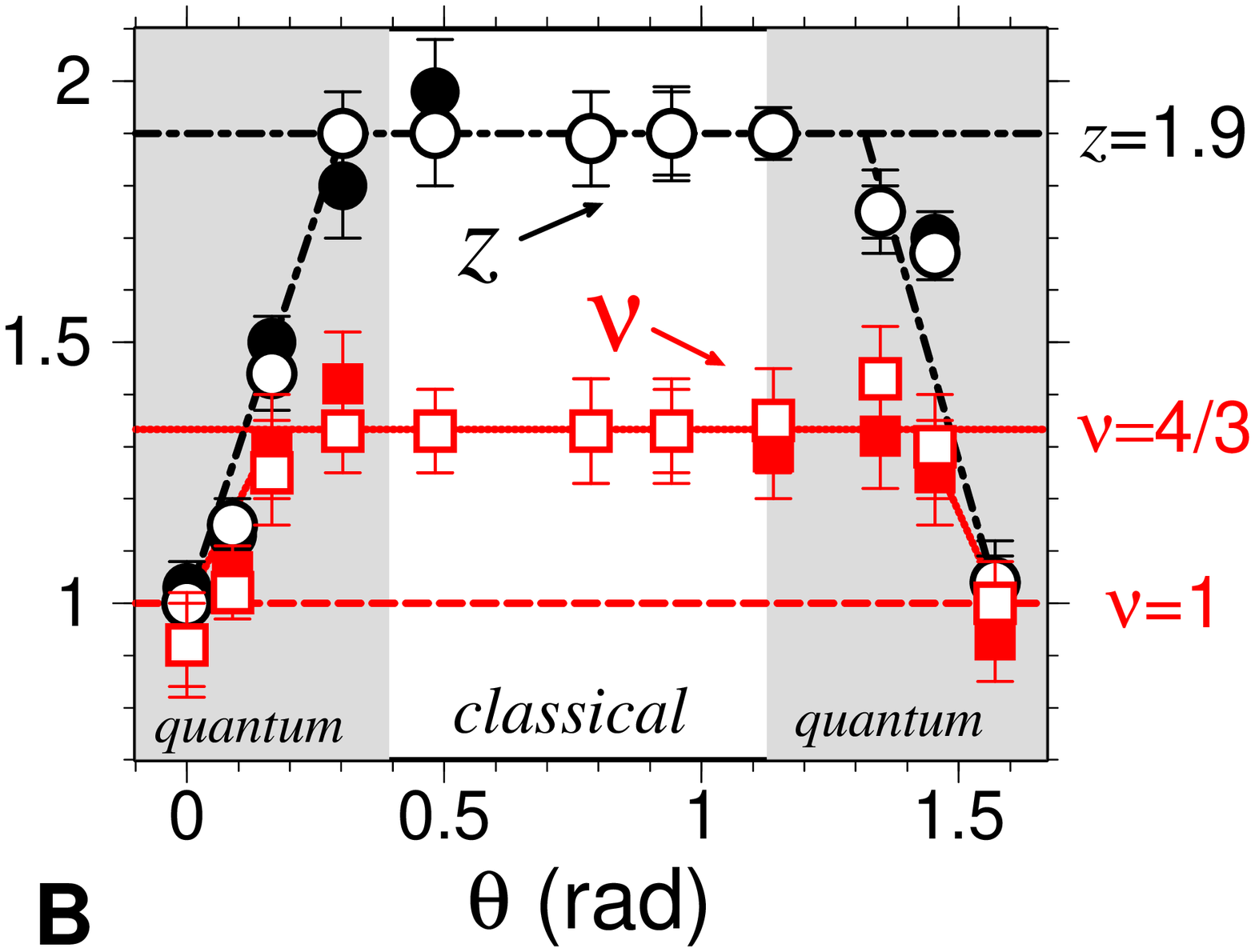}          
\caption{\label{f.exp} (A) Scaling plot of
$\xi_x$ (open symbols), $\xi_y$ (full symbols) for
$P=1$ (randomly coupled dimers). (B) Critical exponents $z$ and $\nu$
for the magnetic transition in the inhomogeneously bond-diluted
S=1/2 Heisenberg antiferromagnet. The angle $\theta$ parametrizes
the critical line in Fig. \ref{f.phdiagr}. Open (full) symbols refer
to estimates obtained through the scaling of $\xi_x$
($\xi_y$). The shaded areas correspond to the regions of parameters
in which the transition is quantum. All lines are guides to the eye.}
\vskip -.2cm
\end{figure}    

 What are the critical exponents associated with 
this quantum phase transition? The finite-temperature 
quantum critical behavior of the correlation length
$\xi \sim T^{-1/z}$ determines the dynamical critical exponent $z$ 
\cite{Sondhietal97,Sachdev99}, whereas 
the critical exponent $\nu$ governs the divergence
of the correlation length at the $T=0$ critical point, 
$\xi \sim |\bm{P}-\bm{P}_c|^{-\nu}$. ${\bm P}_c$ can be 
obtained as the value of disorder at which $\xi$
exhibits a power-law behavior as a function of $T$,
and it is in agreement with the estimates coming from the 
$T=0$ analysis described above. From the scaling relation 
\begin{equation}
\xi({\bm P};T) \sim \frac{1}{T^{1/z}}
 F_{\xi}(T~|{\bm P} -{\bm P}_c|^{\nu z})
\end{equation}
the exponent $\nu$ can be  
determined as the one realizing the best collapse
of $T^{1/z}\xi({\bm P};T)$ curves plotted versus  
$T |{\bm P} -{\bm P}_c|^{\nu z}$ for different $\bm P$
values, as shown in Fig. \ref{f.exp}(A).
The values extracted for $\nu$ and $z$ are shown 
in Fig. \ref{f.exp}(B) as a function of the angle 
$\theta = \tan^{-1}[(1-P')/(1-P)]$ which parametrizes the
critical curve of Fig. \ref{f.phdiagr}.
The $\nu$ exponent is found to satisfy
the Harris criterion for disordered systems 
\cite{Chayesetal86} $\nu \geq 2/d = 1$ within error bars.
Moreover, we observe that $z$ and $\nu$ are almost constant over
a large portion of the critical line, taking the values 
$z \approx 1.9$ and $\nu\approx 4/3$, and they both show 
a drop to $z , \nu\approx 1$ only near the edges of the critical 
line in the strongly quantum regimes 
($P \rightarrow 1$ and $P' \rightarrow 1$).  
Remarkably, such drop happens \emph{away}
from the multicritical points at which the percolation transition 
and the magnetic transition depart from each other.
This implies that, for moderate
quantum fluctuations, the magnetic quantum phase transition 
retains a {\it percolative nature}, in full agreement with
the picture of quantum percolation, in which the
main effect of quantum fluctuations is 
a spatial renormalization of the percolating cluster.
In particular, the correlation length exponent $\nu$
is clearly consistent with the classical value $\nu=4/3$
for two-dimensional percolation \cite{StaufferA94}.
The $z$ exponent can also be related to the critical
exponents of the classical phase transition. 
At and around the homogeneous percolation point $P=P'=0.5$,
where the transition is classical, the exponent $z$
is the scaling dimension of the {\it finite-size} gap
of the percolating cluster diverging 
in the thermodynamic limit, $\Delta \sim L^{-z}$.
Since the number of sites $N_c$ of this cluster diverges as 
$N_c \sim L^D$ with $D=91/48$ \cite{StaufferA94}, if 
the finite-size gap vanishes as $\Delta \sim N_c^{-\rho}$ 
it is easy to see that $z = \rho D$. For clusters
with the geometry of the chain or of the square lattice,
$\rho = 1$ \cite{Whiteetal94,Carlson89}. 
Extending this result to the 
fractal percolating cluster we obtain $z$=$D$=1.89..,
in agreement with the QMC result.

The concept of quantum percolation has previously been 
invoked in the context of weakly 
interacting electronic systems, such as in
metal-to-insulator transitions of granular metals 
\cite{Sheng92}, and in the integer quantum Hall
effect of semiconductor heterostructures \cite{Trugman83}.
In these examples, quantum mechanics enters mostly through
single-particle phenomena, such as 
localization effects of single-electron wave-functions 
and quantum tunneling. In contrast, the quantum percolation 
discussed in the present work occurs in a {\it strongly 
correlated electron system}.
Its quantum aspects reside in the formation of local 
{\it many-body} quantum states, {\it i.e.} dimer singlet 
states or RVB states, and it ultimately has the nature of
a collective quantum phase transition. The specific example
of inhomogeneous bond dilution of the square-lattice Heisenberg 
antiferromagnet offers a remarkable realization
of such scenario, with continuous tuning of the
quantum nature of the percolative transition in the system.
A continuous modulation of quantum effects at a percolation
transition can also be achieved by progressively coupling 
two square-lattice layers which are site-diluted at
exactly the same lattice sites \cite{VajkG02}.
In that case, nonetheless, the estimated critical exponents for
the magnetic quantum phase transition differ substantially
from those of classical percolation and are much closer
to those of a quantum phase transition in a clean 
system, suggesting that quantum effects alter 
the percolative nature of the transition. 

 From the experimental point of view, if different
non-magnetic ions mediate the intra-dimer 
interactions ($b$ bonds of Fig. \ref{structures}(A)) 
with respect to the inter-dimer 
ones ($b'$ bonds of Fig. \ref{structures}(A)) in a square
lattice antiferromagnet, 
selective chemical substitution of these two types 
of ions offers a way to realize 
inhomogeneous bond disorder. Extensions of the
picture given in this paper would be necessary to 
mimic a realistic experimental situation. Nevertheless,
the effects leading to a quantum correction of the 
percolation threshold should be robust, and they 
should persist in presence of more general models 
of inhomogeneous bond disorder. 

We acknowledge fruitful discussions with B. Normand, 
H. Saleur and A. Sandvik. S.H. acknowledges hospitality
at KITP, Santa Barbara. This work is supported by the NSF
through grant No. DMR-0089882. Computational facilities
have been generously provided by the HPCC-USC Center.

\end{document}